\documentclass[11pt]{article}

\usepackage{amsxtra,amssymb,amsthm,amsmath,latexsym}
\usepackage{graphicx}
\usepackage{epsfig}
\usepackage{afterpage}

\newcommand{\R}{{\mathbb R}}

\newcommand{\dl}{{\delta}}
\newcommand{\bee}{\begin{equation*}}
\newcommand{\eee}{\end{equation*}}
\newcommand{\be}{\begin{equation}}
\newcommand{\ee}{\end{equation}}

\title{Scattering of scalar waves by many small particles}
\author{A.G. Ramm \\
\small Department of Mathematics\\[-0.8ex]
\small Kansas State University, Manhattan, KS 66506-2602, USA\\
\small \texttt{ramm@math.ksu.edu}}

\date{}
\begin{document}

\maketitle
\begin{abstract}Formulas are derived for solutions of many-body wave
scattering problems by small particles in the case of acoustically
soft, hard, and impedance particles embedded in an inhomogeneous
medium. The limiting case is considered, when the size $a$ of small
particles tends to zero while their number tends to infinity at a
suitable rate. Equations for the limiting effective
(self-consistent) field in the medium are derived.
\end{abstract}

{\it PACS}: 02.30.Rz; 02.30.Mv; 41.20.Jb

{\it MSC}: \,\, 35Q60;78A40;  78A45; 78A48;

\noindent\textbf{Key words:} wave scattering
by many small bodies; smart materials.
\section{Introduction}
There is a large literature on wave scattering by small bodies,
starting from Rayleigh's work (1871). For the problem of wave
scattering by one body an analytical solution was found only for the
bodies of special shapes, for example, for balls and ellipsoids. If
the scatterer is small then the scattered field can be calculated
analytically for bodies of arbitrary shapes, see \cite{R476}, where
this theory is presented.

The many-body wave scattering problem was discussed in the
literature mostly numerically, if the number of scatterers is small,
or under the assumption that the influence of the waves, scattered
by other particles on a particular particle is negligible ( see
\cite{M}, where one finds a large bibliography, 1386 entries). This
corresponds to the case when the distance $d$ between neighboring
particles is much larger than the wavelength $\lambda$, and the
characteristic size $a$ of a small body (particle) is much smaller than
$\lambda$. By $k=\frac{2\pi}{\lambda}$ the wave number is denoted.

The basic results of our paper consist of: 

i) Derivation of formulas for
the scattering amplitude for the wave scattering problem by one
small ($ka\ll1$) body of an arbitrary shape under the Dirichlet,
impedance, or Neumann boundary condition (acoustically soft,
impedance, or hard particle), 

ii) Solution to many-body wave
scattering problem by many such particles under the assumptions
$a\ll d$ and $a\ll \lambda$, where $d$ is the minimal distance
between neighboring particles, 

iii) Derivation of the equations for
the limiting effective (self-consistent) field in the medium when
$a\to 0$ and the number $M=M(a)$ of the small particles tends to
infinity at an appropriate rate, 

iv) Derivation of linear algebraic
systems for solving many-body wave scattering problems; these system
are not obtained by a discretization of boundary integral equations.

Let us formulate the wave scattering problems we deal with. First,
let us consider a one-body scattering problem. Let $D_1$ be a bounded
domain in $\R^3$ with a sufficiently smooth boundary $S_1$. The
scattering problem consists of finding the solution to the problem:
\be\label{e1} (\nabla^2+k^2)u=0\text{  in  } D'_1:=\R^3\setminus
D_1, \ee 
\be\label{e2} \Gamma u=0\text{  on }S_1, \ee \be\label{e3}
u=u_0+v, \ee where \be\label{e4} u_0=e^{ik\alpha\cdot x}, \quad
\alpha\in S^2, \ee $S^2$ is the unit sphere in $\R^3$, $u_0$ is the
incident field, $v$ is the scattered field satisfying the radiation
condition 
\be\label{e5} v_r-ikv=o\left(\frac{1}{r}\right),\quad
r:=|x|\to \infty,\ v_r:=\frac{\partial v}{\partial r}, \ee $\Gamma
u$ is the boundary condition (bc) of one of the following types
\be\label{e6} \Gamma u=\Gamma_1u=u\quad \text{(Dirichlet bc)}, \ee
\be\label{e7} \Gamma u=\Gamma_2u=u_N-\zeta_1u,\quad \text{
Im}\zeta_1\leq 0,\ \text{( impedance bc)}, \ee where $\zeta_1$ is a
constant, $N$ is the unit normal to $S_1$, pointing out of $D_1$,
and \be\label{e8} \Gamma u=\Gamma_3u=u_N, \ \text{( Neumann bc)}. \ee It 
is well known (see,
e.g., \cite{R190}) that problem \eqref{e1}-\eqref{e3} has a unique
solution. We now assume that \be\label{e9} a:=0.5\,
\text{diam}D_1,\quad ka\ll 1, \ee and look for the solution to problem
\eqref{e1}-\eqref{e3} of the form \be\label{e10} u(x)=u_0(x)+\int_{S_1}
g(x,t)\sigma_1(t)dt,\quad g(x,y):=\frac{e^{ik|x-y|}}{4\pi |x-y|},
\ee where $dt$ is the element of the surface area of $S_1$. One can
prove that the unique solution to the scattering problem
\eqref{e1}-\eqref{e3} with any of the boundary conditions
\eqref{e6}-\eqref{e8} can be found in the form \eqref{e10}, and the
function $\sigma_1$ in equation \eqref{e10} is uniquely defined from the
boundary condition \eqref{e2}. The scattering amplitude
$A(\beta,\alpha)=A(\beta,\alpha,k)$ is defined by the formula
\be\label{e11} v=\frac{e^{ikr}}{r}A(\beta,\alpha,k)+o\left(
\frac{1}{r}\right),\quad r\to \infty,\ \beta:=\frac{x}{r}.\ee The
equations for finding $\sigma_1$ are: \be\label{e3_1}
\int_{S_1}g(s,t)\sigma_1(t)dt=-u_0(s), \ee \be\label{e3_2}
u_{0N}-\zeta_1u_0+\frac{A\sigma_1-\sigma_1}{2}-\zeta_1\int_{S_1}g(s,t)\sigma_1(t)dt=0,
\ee \be\label{e3_3} u_{0N}+\frac{A\sigma_1-\sigma_1}{2}=0, \ee
respectively, for conditions \eqref{e6}-\eqref{e8}. The operator $A$
is defined as follows: \be\label{e3_4}
A\sigma:=2\int_{S_1}\frac{\partial}{\partial N_s}g(s,t)\sigma_1(t)dt.\ee
Equations \eqref{e3_1}-\eqref{e3_3} are uniquely solvable, but there are
no analytic formulas for their solutions for bodies of arbitrary
shapes. However, if the body $D_1$ is small, $ka\ll 1$, one can
rewrite \eqref{e10} as \be\label{e12}
u(x)=u_0(x)+g(x,0)Q_1+\int_{S_1}[g(x,t)-g(x,0)]\sigma_1(t)dt, \ee
where \be\label{e13} Q_1:=\int_{S_1}\sigma_1(t)dt, \ee and $0\in
D_1$ is the origin.

If $ka\ll1$, then we prove that \be\label{e14} |g(x,0)Q_1|\gg
\left|\int_{S_1}[g(x,t)-g(x,0)]\sigma_1(t)dt\right|,\quad |x|>a. \ee
Therefore, the scattered field is determined outside $D_1$
by a single number $Q_1$.
This number can be obtained analytically
without solving equations \eqref{e3_1}-\eqref{e3_2}. The case
\eqref{e3_3} requires a special approach by the reason discussed in
detail later. 

Let us give the results for equations \eqref{e3_1} and
\eqref{e3_2} first. For equation \eqref{e3_1} one has \be\label{e15}
Q_1=\int_{S_1}\sigma_1(t)dt=-Cu_0(0)[1+o(1)],\quad a\to 0, \ee where
$C$ is the electric capacitance of a perfect conductor with the
shape $D_1$. For equation \eqref{e3_2} one has \be\label{e16}
Q_1=-\zeta |S_1|u_0(0)[1+o(1)],\quad a\to 0, \ee where $|S_1|$ is
the surface area of $S_1$. The scattering amplitude for problem
\eqref{e1}-\eqref{e3} with $\Gamma=\Gamma_1$ (acoustically soft
particle) is \be\label{e17}
A_1(\beta,\alpha)=-\frac{C}{4\pi}[1+o(1)], \ee since
$$u_0(0)=e^{ik\alpha\cdot x}|_{x=0}=1.$$ {\it Therefore,  in this case 
the
scattering is isotropic and of the order $O(a)$, because the
capacitance $C=O(a)$. }

The scattering amplitude for problem
\eqref{e1}-\eqref{e3} with $\Gamma=\Gamma_2$ (small impedance
particles) is : \be\label{e18}
A_2(\alpha,\beta)=-\frac{\zeta_1|S_1|}{4\pi}[1+o(1)], \ee since
$u_0(0)=1$. 

{\it In this case the scattering is also isotropic, and of
the order $O(\zeta|S_1|)$.} 

If $\zeta_1=O(1)$, then $A_2=O(a^2)$,
because $|S_1|=O(a^2)$. If
$\zeta_1=O\left(\frac{1}{a^\kappa}\right)$, $\kappa\in(0,1)$, then
 $A_2=O(a^{2-\kappa})$. The case $\kappa=1$ was considered in \cite{R509}.

The scattering amplitude for problem
\eqref{e1}-\eqref{e3} with $\Gamma=\Gamma_3$ (acoustically hard
particles) is \be\label{e19}
A_3(\beta,\alpha)=-\frac{k^2|D_1|}{4\pi}(1+\beta_{pq}\beta_p\alpha_q),\,\,\text{
if } u_0=e^{ik\alpha\cdot x}. \ee 
Here and below summation is understood over
the repeated indices, $\alpha_q=\alpha \cdot e_q$, $\alpha \cdot e_q$ 
denotes the 
dot product of two vectors in $\R^3$, $p,q=1,2,3$,
$\{e_p\}$ is an orthonormal Cartesian basis of $\R^3$, 
$|D_1|$ is the volume of $D_1$, $\beta_{pq}$
is the magnetic polarizability tensor defined as follows
(\cite{R476}, p.62): \be\label{e20}
\beta_{pq}:=\frac{1}{|D_1|}\int_{S_1}t_p\sigma_{1q}(t)dt, \ee 
 $\sigma_{1q}$ is the solution to
the equation \be \label{e21} \sigma_{1q}(s)=A_0\sigma_{1q}-2N_q(s),
\ee  $N_q(s)=N(s)\cdot e_q$, $N=N(s)$ is the unit outer normal to $S_1$ at 
the point $s$, i.e., the normal pointing out of $D_1$, and $A_0$
is the operator $A$ at $k=0$. For small bodies $\|A-A_0\|=o(ka)$.

If $u_0(x)$ is an arbitrary field satisfying equation \eqref{e1},
not necessarily the plane wave $e^{ik\alpha\cdot x}$, then
\be\label{e22}
A_3(\beta,\alpha)=\frac{|D_1|}{4\pi}\left(ik\beta_{pq}\frac{\partial
u_0}{\partial x_q} \beta_p+\triangle u_0\right). \ee The
above formulas are derived in Section 2. In Section 3 we develop a
theory for many-body wave scattering problem and derive the
equations for effective field in the medium, in which many small
particles are embedded, as $a\to 0$.

The results, presented in this paper, are based on the earlier works of 
the author (\cite{R509}-\cite{R597}). Our presentation and some of the 
results are novel. These results and methods of their derivation 
differ much from those in the
homogenization theory (\cite{JKO}, \cite{MK}). The
differences are: 

i) no periodic structure in the problems is assumed,

ii) the operators in our problems are non-selfadjoint and have
continuous spectrum, 

iii) the limiting medium is not homogeneous and its
parameters are not periodic, 

iv) the technique for passing to the limit
is different from one used in homogenization theory.

\section{Derivation of the formulas for one-body wave scattering
problems}

Let us recall the known result (see e.g., \cite{R190})
\be\label{e23} \frac{\partial }{\partial
N_s^-}\int_{S_1}g(x,t)\sigma_1(t)dt=\frac{A\sigma_1-\sigma_1}{2}\ee
concerning the limiting value of the normal derivative of single-layer
potential from outside. Let $x_m\in D_m$, $t\in S_m$, $S_m$ is the
surface of $D_m$, $a=0.5\,\text{diam} D_m.$ 

In this Section $m=1$, and $x_m=0$ is the origin.

We assume  that $ka\ll 1$, $ad^{-1}\ll 1$,
so $|x-x_m|=d\gg a$. Then \be\label{e24}
\frac{e^{ik|x-t|}}{4\pi|x-t|}=\frac{e^{ik|x-x_m|}}{4\pi|x-x_m|}
e^{-ik(x-x_m)^o\cdot(t-x_m)}\left(1+O(ka+\frac{a}{d})\right),
\ee \be\label{e25}
k|x-t|=k|x-x_m|-k(x-x_m)^o\cdot(t-x_m)+O\left(\frac{ka^2}{d}\right),
\ee where 
$$d=|x-x_m|,\quad (x-x_m)^o:=\frac{x-x_m}{|x-x_m|},$$ and
\be\label{e26} \frac{|x-t|}{|x-x_m|}=1+O\left(\frac{a}{d}\right).
\ee 
Let us  derive estimate \eqref{e15}. Since
$|t|\leq a$ on $S_1$, one has 
$$g(s,t)=g_0(s,t)(1+O(ka)),$$ where
$g_0(s,t)=\frac{1}{4\pi|s-t|}$. Since $u_0(s)$ is a smooth function,
one has $|u_0(s)-u_0(0)|=O(a)$. Consequently, equation \eqref{e3_1}
can be considered as an equation for electrostatic charge
distribution $\sigma_1(t)$ on the surface $S_1$ of a perfect
conductor $D_1$, charged to the constant potential $-u_0(0)$ (up to
a small term of the order $O(ka)$). It is known that the total charge
$Q_1=\int_{S_1}\sigma_1(t)dt$ of this conductor is equal to
\be\label{e27} Q_1=-Cu_0(0)(1+O(ka)), \ee where $C$ is the electric
capacitance of the perfect conductor with the shape $D_1$. 

Analytic
formulas for electric capacitance $C$ of a perfect conductor of an
arbitrary shape, which allow to calculate $C$ with a desired
accuracy, are derived in \cite{R476}. For example, the zeroth
approximation formula is \be\label{e28}
C^{(0)}=\frac{4\pi|S_1|^2}{\int_{S_1}\int_{S_1}\frac{dsdt}{r_{st}}},\quad
r_{st}=|t-s|, \ee and we assume in \eqref{e28} that $\epsilon_0=1$,
where $\epsilon_0$ is the dielectric constant of the 
homogeneous medium in which the 
perfect conductor is placed. Formula
\eqref{e27} is formula \eqref{e15}. If $u_0(x)=e^{ik\alpha\cdot x}$,
then $u_0(0)=1$, and $Q_1=-C(1+O(ka))$. In this case
$$A_1(\beta,\alpha)=\frac{Q_1}{4\pi}=-\frac{C}{4\pi}[1+O(ka)],$$ which
is formula \eqref{e17}.

{\it Consider now wave scattering by an impedance particle}.

Let us derive formula \eqref{e16}. Integrate equation \eqref{e3_2}
over $S_1$, use the divergence formula \be\label{e29}
\int_{S_1}u_{0N} ds=\int_{D_1}\nabla^2u_0 dx=-k^2\int_{D_1}u_0
dx=k^2|D_1|u_0(0)[1+o(1)], \ee where $|D_1|=O(a^3)$, and the formula
\be\label{e30} -\zeta_1\int_{S_1}u_0ds=-\zeta_1|S_1|u_0(0)[1+o(1)].
\ee Futhermore $|\int_{S_1}g(s,t)ds|=O(a),$ so \be\label{e31}
\zeta_1\int_{S_1}ds\int_{S_1}g(s,t)\sigma_1(t)dt=O(aQ_1). \ee
Therefore, the term \eqref{e31} is negligible compared with $Q_1$
as  $a\to 0$. Finally, if $ka\ll1$, then
$g(s,t)=g_0(s,t)\left( 1+ik|s-t|+\hdots \right),$ and 
\be\label{e32}
\frac{\partial}{\partial N_s}g(s,t)=\frac{\partial }{\partial
N_s}g_0(s,t)[1+O(ka)]. \ee Denote by $A_0$ the operator
\be\label{e33} A_0\sigma=2\int_{S_1}\frac{\partial
g_0(s,t)}{\partial N_s}\sigma_1(t)dt. \ee It is known from the
potential theory that 
\be\label{e34} \int_{S_1}A_0\sigma_1
ds=-\int_{S_1}\sigma_1(t)dt,\quad 2\int_{S_1} \frac{\partial
g_0(s,t)}{\partial N_s}ds=-1,\quad t\in S_1.\ee Therefore,
\be\label{e35}
\int_{S_1}ds\frac{A\sigma_1-\sigma_1}{2}=-Q_1[1+O(ka)]. \ee
Consequently, from fromulas \eqref{e29}-\eqref{e35} one gets formula
\eqref{e18}.

One can see that the wave scattering by an impedance particle is
isotropic, and the scattered field is of the order
$O(\zeta_1|S_1|)$. Since $|S_1|=O(a^2)$, one would have
$O(\zeta_1|S_1|)=O(a^{2-\kappa})$ if
$\zeta_1=O\left(\frac{1}{a^\kappa}\right)$, $\kappa\in(0,1)$.

{\it Consider now wave scattering by an acoustically hard small particle,
i.e., the problem with the Neumann boundary condition.} 

In this case 
we will prove that: 

i)  The scattering is
anisotropic, 

ii) It is defined not by a single number, as in the previous
two cases, but by a tensor, 

and 

iii) The order of the scattered field is
$O(a^3)$ as $a\to 0$, for a fixed $k>0$, i.e., the scattered field
is much smaller than in the previous two cases. 

When one integrates
over $S_1$ equation \eqref{e3_2}, one gets \be\label{e36}
Q_1=\int_{D_1}\nabla^2u_0 dx=\nabla^2u_0(0)|D_1|[1+o(1)],\quad a\to
0. \ee Thus, $Q_1=O(a^3)$. Therefore, the contribution of the term
$e^{-ikx^o\cdot t}$ in formula \eqref{e24} with $x_m=0$ will be also
of the order $O(a^3)$ and should be taken into account, {\it in contrast
to the previous two cases.} Namely, \be\label{e37}
u(x)=u_0(x)+g(x,0)\int_{S_1}e^{-ik\beta\cdot t}\sigma_1(t)dt,\quad
\beta:=\frac{x}{|x|}=x^o. \ee One has \be\label{e38}
\int_{S_1}e^{-ik\beta\cdot
t}\sigma_1(t)dt=Q_1-ik\beta_p\int_{S_1}t_p\sigma_1(t)dt, \ee where
the terms of higher order of smallness are neglected and
summation over index $p$ is understood. The function $\sigma_1$
solves equation \eqref{e3_3}: \be\label{e39}
\sigma_1=A\sigma_1+2u_{0N}=A\sigma_1+2ik\alpha_qN_qu_0(s),\quad s\in
S_1 \ee if $u_0(x)=e^{ik\alpha\cdot x}$. 

Comparing \eqref{e39} with
\eqref{e21}, using \eqref{e20}, and taking into account that $ka\ll
1$, one gets \be\label{e40}\begin{split}
-ik\beta_p\int_{S_1}t_p\sigma_1(t)dt&=-ik\beta_p|D_1|\beta_{pq}(-ik\alpha_q)
u_0(0)[1+O(ka)]\\
&=-k^2|D_1|\beta_{pq}\beta_p\alpha_qu_0(0)[1+O(ka)].
\end{split}\ee
From \eqref{e36}, \eqref{e38} and \eqref{e40} one gets formula
\eqref{e19}, because $\nabla^2u_0=-k^2u_0.$ 

If $u_0(x)$ is an
arbitrary function, satisfying equation \eqref{e1}, then
$ik\alpha_q$ in \eqref{e39} is replaced by $\frac{\partial
u_0}{\partial x_q}$, and $-k^2u_0=\triangle u_0$, which yields
formula \eqref{e22}.

This completes the derivation of the formulas for the solution of
scalar wave scattering problem by one small body on the boundary of
which the Dirichlet, or the impedance, or the Neumann boundary
condition is imposed.

\section{Many-body scattering problem}
In this Section we assume that there are $M=M(a)$ small bodies
(particles) $D_m$, $1\leq m\leq M$, $a=0.5\max\text{diam}D_m$, $ka\ll1$. 
The
distance $d=d(a)$ between neighboring bodies is much larger than
$a$, $d\gg a$, but we do not assume that $d\gg \lambda$, so {\it there
may be many small particles on the distances of the order of the
wavelength $\lambda$.} This means that our medium with the embedded
particles is not necessarily diluted. 

We assume that the small bodies are embedded in an arbitrary large but 
finite domain $D$, $D\subset \R^3$, so 
$D_m\subset D$. Denote
$D':=\R^3\setminus D$ and $\Omega:=\cup_{m=1}^M D_m,$ 
$S_m:=\partial
D_m$, $\partial \Omega=\cup_{m=1}^M S_m$. By $N$ we denote a unit
normal to $\partial \Omega$, pointing out of $\Omega$, by $|D_m|$
the volume of the body $D_m$ is denoted. 

The
scattering problem consists of finding the solution to the following
problem \be\label{e41} (\nabla^2+k^2)u=0\text{  in  } \R^3\setminus
\Omega, \ee \be\label{e42} \Gamma u=0\text{   on  } \partial \Omega,
\ee \be\label{e43} u=u_0+v, \ee where $u_0$ is the incident field,
satisfying equation \eqref{e41} in $\R^3$, for example,
$u_0=e^{ik\alpha \cdot x}$, $\alpha\in S^2$, and $v$ is the
scattered field, satisfying the radiation condition \eqref{e5}. The
boundary condition \eqref{e42} can be of the types
\eqref{e6}-\eqref{e8}.

In the case of impedance boundary condition \eqref{e7} we assume
that \be\label{e44} u_N=\zeta_mu\text{ on } S_m,\quad 1\leq m\leq
M,\ee so the impedance may vary from one particle to another. We
assume that \be\label{e45} \zeta_m=\frac{h(x_m)}{a^\kappa},\quad
\kappa\in(0,1), \ee where $x_m\in D_m$ is a point in $D_m$, and
$h(x),$ $x\in D$, is a given function, which we can choose as we
wish, subject to the condition  Im$h(x)\leq 0$. For simplicity we 
assume that $h(x)$ is a
continuous function. 

Let us make the following assumption about the
distribution of small particles: if $\Delta\subset D$ is an
arbitrary open subset of $D$, then the number $\mathcal{N}(\Delta)$
of small particles in $\Delta$, assuming the impedance boundary condition,
is: \be\label{e46}
\mathcal{N}_\zeta(\Delta)=\frac{1}{a^{2-\kappa}}\int_{\Delta}N(x)dx[1+o(1)],
\quad a\to 0, \ee where $N(x)\geq 0$ is a given function.
If the
Dirichlet boundary condition is assumed, then \be\label{e47}
\mathcal{N}_D(\Delta)=\frac{1}{a}\int_{\Delta}N(x)dx[1+o(1)],\quad
a\to 0.\ee The case of the Neumann boundary condition will be
considered later. 

We look for the solution to problem
\eqref{e41}-\eqref{e43} with the Dirichlet boundary condition of the
form \be\label{e48}
u=u_0+\sum_{m=1}^M\int_{S_m}g(x,t)\sigma_m(t)dt,\ee where
$\sigma_m(t)$ are some functions to be determined from the boundary
condition \eqref{e42}. It is proved in \cite{R509} that problem
\eqref{e41}-\eqref{e43} has a unique solution of the form
\eqref{e48}. For any $\sigma_m(t)$ function \eqref{e48} solves
equation \eqref{e41} and satisfies condition \eqref{e43}. The
boundary condition \eqref{e42} determines $\sigma_m$ uniquely.
However, if $M\gg1$, then numerical solution of the system of
integral equations for $\sigma_m$, $1\leq m\leq M$, which one gets
from the boundary condition \eqref{e42}, is practically not
feasible. 

{\it To avoid this principal difficulty we prove that the
solution to scattering problem \eqref{e41}-\eqref{e43} is determined
by $M$ numbers \be\label{e49} Q_m:=\int_{S_m}\sigma_m(t)dt, \ee
rather than $M$ functions $\sigma_m(t)$.} 

This is possible to prove 
if the
particles $D_m$ are small. We derive analytical formulas for $Q_m$
as $a\to 0$. 

Let us define the effective (self-consistent) field
$u_e(x)=u_e^{(j)}(x)$, acting on the $j-$th particle, by the formula
\be\label{e50} u_e(x):=u(x)-\int_{S_j}g(x,t)\sigma_j(t)dt,\quad
|x-x_j|\sim a.\ee Physically this field acts on the $j-$th particle
and is a sum of the incident field and the fields acting from all
other particles: \be\label{e51}
u_e(x)=u_e^{(j)}(x):=u_0(x)+\sum_{m\neq
j}\int_{S_m}g(x,t)\sigma_m(t)dt. \ee Let us rewrite \eqref{e51} as
follows: \be\label{e52} u_e(x)=u_0(x)+\sum_{m\neq j}^M
g(x,x_m)Q_m+\sum_{m\neq
j}^M\int_{S_m}[g(x,t)-g(x,x_m)]\sigma_m(t)dt. \ee 
We want to prove
that the last sum is negligible compared with the first one as $a\to
0$. To prove this, let us give some estimates. One has $|t-x_m|\leq a$, 
$d=|x-x_m|$, \be\label{e53}
|g(x,t)-g(x,x_m)|=\max\left\{
O\left(\frac{a}{d^2}\right),O\left(\frac{ka}{d}\right)\right\},\quad
|g(x,x_m)|=O(1/d). \ee Therefore, if $|x-x_j|=O(a)$, then
\be\label{e54}
\frac{\left|\int_{S_m}[g(x,t)-g(x,x_m)]\sigma_m(t)dt\right|}{|g(x,x_m)Q_m|}\leq
O(ad^{-1}+ka). \ee One can also prove that \be\label{e55}
J_1/J_2=O(ka+ad^{-1}),\ee where $J_1$ is the first sum in
\eqref{e52} and $J_2$ is the second sum in \eqref{e52}. Therefore,
at any point $x\in \Omega'=\R^3\setminus \Omega$ one has
\be\label{e56} u_e(x)=u_0(x)+\sum_{m=1}^M g(x,x_m)Q_m,\quad x\in
\Omega', \ee
where the terms of higher order of smallness are omitted.

\subsection{ The case of acoustically soft particles}
If \eqref{e42} is the Dirichlet condition, then, as we have proved
in Section 2  (see formula \eqref{e27}), one has
\be\label{e57} Q_m=-C_mu_e(x_m). \ee Thus, \be\label{e58}
u_e(x)=u_0(x)-\sum_{m=1}^M g(x,x_m)C_mu_e(x_m), \quad x\in \Omega'.\ee 
One has \be\label{e59}
u(x)=u_e(x)+o(1),\quad a\to 0, \ee so the full field and
effective field are practically the same. 

Let us write a linear
algebraic system (LAS) for finding unknown quantities $u_e(x_m)$:
\be\label{e60} u_e(x_j)=u_0(x_j)-\sum_{m\neq
j}^Mg(x_j,x_m)C_mu_e(x_m). \ee If $M$ is not very large, say
$M=O(10^3)$, then LAS \eqref{e60} can be solved numerically, and 
formula \eqref{e58} can be used for calculation of $u_e(x)$.

Consider the limiting case, when $a\to 0$. One can rewrite
\eqref{e60} as follows: \be\label{e61}
u_e(\xi_q)=u_0(\xi_q)-\sum_{p\neq q}^P
g(\xi_q,\xi_p)u_e(\xi_p)\sum_{x_m\in \Delta_p}C_m, \ee where
$\{\Delta_p\}_{p=1}^P$ is a union of cubes which forms a covering of
$D$, 
$$\max_p diam \Delta_p:=b=b(a)\gg a,$$
 \be\label{e62}
\lim_{a\to 0}b(a)=0. \ee By $|\Delta_p|$ we denote the volume
(measure) of $\Delta_p$, and $\xi_p$ is the center of $\Delta_p$,
or a point $x_p$ in an arbitrary small body $D_p$, located in $\Delta_p$.
Let us assume that there exists the limit \be\label{e63} \lim_{a\to
0}\frac{\sum_{x_m\in \Delta_p}C_m}{|\Delta_p|}=C(\xi_p),\quad
\xi_p\in \Delta_p. \ee 
For example, one may have  \be\label{e64} C_m=c(\xi_p)a
\ee for all $m$ such that $x_m\in \Delta_p$, where $c(x)$ is some
function in $D$. If all $D_m$ are balls of radius $a$,
then $c(x)=4\pi$. We have \be\label{e65}
\sum_{x_m\in\Delta_p}C_m=C_pa\mathcal{N}(\Delta_p)=C_pN(\xi_p)|\Delta_p|[1+o(1)],\quad
a\to 0, \ee so limit \eqref{e63} exists, and \be\label{e66}
C(\xi_p)=c(\xi_p)N(\xi_p).\ee From \eqref{e61},
\eqref{e64}-\eqref{e66} one gets \be\label{e67}
u_e(\xi_q)=u_0(\xi_q)-\sum_{p\neq
q}g(\xi_q,\xi_p)c(\xi_p)N(\xi_p)u_e(\xi_p)|\Delta_p|,\quad 1\leq
p\leq P. \ee Linear algebraic system  \eqref{e67} can be considered as the 
{\it collocation
method for solving integral equation} \be\label{e68}
u(x)=u_0(x)-\int_Dg(x,y)c(y)N(y)u(y)dy. \ee It is proved in
\cite{R573} that system \eqref{e67} is uniquely solvable for all
sufficiently small $b(a)$, and the function \be\label{e69}
u_P(x):=\sum_{p=1}^P\chi_p(x)u_e(\xi_p)\ee converges in $L^\infty
(D)$ to the unique solution of equation \eqref{e68}. 
The function $\chi_p(x)$ in \eqref{e69} is the characteristic function
of the cube $\Delta_p$: it is equal to $1$ in $\Delta_p$ and vanishes 
outside $\Delta_p$.
Thus, if $a\to
0$, the solution to the many-body wave scattering problem in the
case of the Dirichlet boundary condition is well approximated by the
unique solution of the integral equation \eqref{e68}. 

Applying the
operator $L_0:=\nabla^2+k^2$ to \eqref{e68}, and using the formula
$L_0g(x,y)=-\delta(x-y)$, where $\dl(x)$ is the delta-function, one
gets \be\label{e70}\nabla^2u+k^2u-q(x)u=0 \text{ in } \R^3,\quad
q(x):=c(x)N(x).\ee 
The physical conclusion is: 

{\it If one  embeds $M(a)=O(1/a)$ small 
acoustically
soft particles, which are distributed as in \eqref{e47}, then one   
creats, as $a\to 0$, a
limiting medium, which is inhomogeneous,  and has a refraction coefficient
$n^2(x)=1-k^{-2}q(x).$ }

It is interesting from the physical point of
view to note that {\it the limit, as $a\to 0$, of the total volume of the
embedded particles is zero.} 

Indeed, the volume of one particle is
$O(a^3)$, the total number $M$ of the embedded particles is
$O(a^3M)=O(a^2)$, and $\lim_{a\to 0}O(a^2)=0$. 

The second
observation is: if \eqref{e47} holds, then on a unit length
straight line there are $O(\frac{1}{a^{1/3}})$ particles, so the
distance between neighboring particles is $d=O(a^{1/3})$. If
$d=O(a^\gamma)$ with $\gamma>\frac{1}{3}$, then the number of the
embedded particles in a subdomain $\Delta_p$ is
$O(\frac{1}{d^3})=O(a^{-3\gamma})$. In this case, for $3\gamma>1$,
the limit in \eqref{e65} is $C(\xi_p)=\lim_{a\to
0}c_paO(a^{-3\gamma})=\infty$. Therefore, the product of this limit
by $u$ remains finite only if $u=0$ in $D$. Physically this means
that if the distances between neighboring perfectly soft particles
are smaller than $O(a^{1/3})$, namely, they are $O(a^\gamma)$ with
any $\gamma>\frac{1}{3}$, then $u=0$ in $D$. 

On the other hand, if
$\gamma<\frac{1}{3}$, then the limit $C(\xi_p)=0$, and $u=u_0$ in
$D$, so that the embedded particles do not change, in the limit
$a\to 0$, properties of the medium. 

This concludes our discussion of
the scattering problem for many acoustically soft particles.

\subsection{ Wave scattering by many impedance particles}
We assume now that \eqref{e45} and \eqref{e46} hold, use the exact
boundary condition \eqref{e42} with $\Gamma=\Gamma_2$, that is,
\be\label{e71}
u_{eN}-\zeta_mu_e+\frac{A_m\sigma_m-\sigma_m}{2}-\zeta_m\int_{S_m}g(s,t)\sigma_m(t)dt=0,
\ee and integrate \eqref{e71} over $S_m$ in order to derive an
analytical asymptotic formula for $Q_m=\int_{S_m}\sigma_m(t)dt.$ 

We have 
\be\label{e72} \int_{S_m}u_{eN}ds=\int_{D_m}\nabla^2 u_e
dx=O(a^3), \ee \be\label{e73}
\int_{S_m}\zeta_mu_e(s)ds=h(x_m)a^{-\kappa}|S_m|u_e(x_m)[1+o(1)],\quad
a\to 0, \ee \be\label{e74}
\int_{S_m}\frac{A_m\sigma_m-\sigma_m}{2}ds=-Q_m[1+o(1)],\quad a\to
0, \ee and \be\label{e75}
\zeta_m\int_{S_m}\int_{S_m}g(s,t)\sigma_m(t)dt=h(x_m)a^{1-\kappa}Q_m=o(Q_m),\quad
0<\kappa<1. \ee From \eqref{e71}-\eqref{e75} one finds
\be\label{e76} Q_m=-h(x_m)a^{2-\kappa}|S_m|a^{-2}u_e(x_m)[1+o(1)].
\ee This yields the formula for the approximate solution to the
wave scattering problem for many impedance particles:
\be\label{e77}
u(x)=u_0(x)-a^{2-\kappa}\sum_{m=1}^Mg(x,x_m)b_mh(x_m)u_e(x_m)[1+o(1)],
\ee where $$b_m:=|S_m|a^{-2}$$ are some positive numbers which depend
on the geometry of $S_m$ and are independent of $a$. For example, if
all $D_m$ are balls of radius $a$, then $b_m=4\pi$.

{\it A linear algebraic system for $u_e(x_m)$}, analogous to \eqref{e60},
is \be\label{e78} u_e(x_j)=u_0(x_j)-a^{2-\kappa}\sum_{m=1,m\neq j}^M
g(x_j,x_m)b_m h(x_m)u_e(x_m). \ee The integral equation for the
limiting effective field in the medium with embedded small
particles, as $a\to 0$, is \be\label{e79}
u(x)=u_0(x)-b\int_{D}g(x,y)N(y)h(y)u(y)dy, \ee where \be\label{e80}
u(x)=\lim_{a\to 0}u_e(x),\ee and we have assumed in \eqref{e79} for
simplicity that $b_m=b$ for all $m$, that is, all small particles
are of the same shape and size. 

Applying operator $L_0=\nabla^2+k^2$
to equation \eqref{e79}, one finds the differential equation for the
limiting effective field $u(x)$: \be\label{e81}
(\nabla^2+k^2-bN(x)h(x))u=0\text{  in }\R^3, \ee and $u$ satisfies
condition \eqref{e43}. 

{\it The conclusion is: the limiting medium is
inhomogeneous, and its properties are described by the function
\be\label{e82} q(x):=bN(x)h(x). \ee }
Since the choice of the
functions $N(x)\geq 0$ and $h(x)$, Im$h(x)\leq 0$, is at our
disposal, we can create the medium with desired properties by
embedding many small impedance particles, with suitable impedances,
according to the distribution law \eqref{e46} with a suitable
$N(x)$. The function \be\label{e83} 1-k^{-2}q(x)=n^2(x) \ee is the
refraction coefficient of the limiting medium. Given a desired
refraction coefficient $n^2(x)$, Im$n^2(x)\geq 0$, one can find
$N(x)$ and $h(x)$ so that \eqref{e83} holds, that is, one can create
a material with a desired refraction coefficient by embedding into a
given material many small particles with suitable boundary
impedances.

This concludes our discussion of the wave scattering
problem with many small impedance particles.
\subsection{Wave scattering by many  acoustically hard particles}
Consider now the case of acoustically hard particles, i.e., the case
of Neumann boundary condition. The exact boundary integral equation
for the function $\sigma_m$
in this case is: \be\label{e84}
u_{eN}+\frac{A_m\sigma_m-\sigma_m}{2}=0. \ee Arguing as in Section
2, see formulas \eqref{e36}-\eqref{e40}, one obtains \be\label{e85}
u_e(x)=u_0(x)+\sum_{m=1}^Mg(x,x_m)\left[\triangle
u_e(x_m)+ik\beta_{pq}^{(m)}\frac{(x_p-(x_m)_p)}{|x-x_m|}\frac{\partial
u_e(x_m)}{\partial (x)_q}\right]|D_m|. \ee Here we took into account
that the unit vector $\beta$ in \eqref{e40} is now the vector
$\frac{x-x_m}{|x-x_m|}$, and
$\beta_p=\frac{(x)_p-(x_m)_p}{|x-x_m|}$, where $(x)_p:=x\cdot e_p$
is the $p-$th component of vector $x$ in the Euclidean orthonormal basis
$\{e_p\}_{p=1}^3$. 

There are three sets of unknowns in \eqref{e85}:
$u_e(x_m)$, $\frac{\partial u_e(x_m)}{\partial (x)_q}$, and
$\triangle u_e(x_m)$, $1\leq m\leq M$, $1\leq q\leq 3$. To obtain
linear algebraic system for $u_e(x_m)$ and $\frac{\partial
u_e(x_m)}{\partial (x)_q}$ one sets $x=x_j$ in \eqref{e85}, takes
the sum in \eqref{e85} with $m\neq j$. This yields the first set of
equations for finding these unknowns. Then one takes derivative 
of equation \eqref{e85} with respect to $(x)_q$, sets $x=x_j$,
and takes the sum in \eqref{e85} with $m\neq j$. This yields the second 
set of equations for finding these unknowns. Finally, one takes 
Laplacian of equation \eqref{e85}, sets $x=x_j$, and takes the sum in 
\eqref{e85} with $m\neq j$. This yields the third set of linear 
algebraic equations
for finding $u_e(x_m)$, $\frac{\partial u_e(x_m)}{\partial (x)_q}$,
and $\Delta u_e(x_m)$. 

Passing to the limit $a\to 0$
in equation \eqref{e85}, yields the equation for the limiting field
\be\label{e86}
u(x)=u_0(x)+\int_Dg(x,y)\left(\rho(y)\nabla^2u(y)+ik\frac{\partial
u (y)}{\partial y_q}\frac{x_p-y_p}{|x-y|}B_{pq}(y)\right)dy, \ee where
$\rho(y)$ and $B_{pq}(y)$ are defined below, see formulas \eqref{e88}
and \eqref{e89}. 

Let us derive equation \eqref{e86}.
We start by transforming the sum in \eqref{e85}. Let
$\{\Delta_l\}_{l=1}^L$ be a covering of $D$ by cubes $\Delta_l$,
$\max_l$ diam$\Delta_l=b=b(a)$. We assume that 
$$b(a)\gg d\gg a, \qquad \lim_{a\to
0}b(a)=0.$$ 
Thus, there are many small particles $D_m$ in $\Delta_l$.
Let $x_l$ be a point in $\Delta_l$. One has
\be\label{e87}\begin{split} &\sum_{m=1}^Mg(x,x_m)\left[\triangle
u_e(x_m)+ik\frac{\partial u_e(x_m)}{\partial
(x)_q}\beta_{pq}^{(m)}\frac{((x)_p-(x_m)_p)}{|x-x_m|}\right]|D_m|\\
&=\sum_{l=1}^Lg(x,x_l)\left[\triangle u_e(x_l)\sum_{x_m\in
\Delta_l}|D_m|+ik\frac{\partial u_e(x_l)}{\partial
(x)_q}\frac{((x)_p-(x_l)_p)}{|x-x_l|}\sum_{x_m\in
\Delta_l}\beta_{pq}^{(m)}|D_m|\right].\end{split}\ee Assume that the
following limit exist: \be\label{e88} \lim_{a\to 0,y\in
\Delta_l}\frac{\sum_{x_m\in \Delta_l} |D_m|}{|\Delta_l|}=\rho(y),
\ee \be\label{e89} \lim_{a\to 0,y\in \Delta_l}\frac{\sum_{x_m\in
\Delta_l} \beta_{pq}^{(m)}|D_m|}{|\Delta_l|}=B_{pq}(y), \ee and
\be\label{e90} \lim_{a\to 0}u_e(y)=u(y),\quad \lim_{a\to
0}\frac{\partial u_e(y)}{\partial (y)_q}=\frac{\partial u (y)}{\partial
y_q},\quad \lim_{a\to 0}\nabla^2 u_e(y)=\nabla^2 u(y). \ee
 Then, the sum in \eqref{e87} converges to
\be\label{e91} \int_Dg(x,y)\left(\rho(y)\nabla^2u(y)+ik\frac{\partial
u(y)}{\partial y_q}\frac{x_p-y_p}{|x-y|}B_{pq}(y)\right)dy. \ee
Consequently, \eqref{e85} yields in the limit $a\to 0$ equation
\eqref{e86}. Equation \eqref{e86} cannot be reduced to a
differential equation for $u(x)$, because \eqref{e86} is an
integrodifferential equation whose integrand depends on $x$ and $y$.

\section{Scattering by small particles embedded in an inhomogeneous
medium} 
Suppose that the operator $\nabla^2+k^2$ in \eqref{e1} and
in \eqref{e41} is replaced by the operator
$L_0=\nabla^2+k^2n_0^2(x)$, where $n_0^2(x)$ is a known function,
\be\label{e92} \text{Im}\,n_0^2(x)\geq 0. \ee The function $n_0^2(x)$
is the refraction coefficient of an inhomogeneous medium in which
many small particles are embedded. The results, presented in Section
1-3 remain valid if one replaces function $g(x,y)$ by the Green's
function $G(x,y)$, \be\label{e93}
[\nabla^2+k^2n_0^2(x)]G(x,y)=-\delta(x-y),\ee satisfying the
radiation condition. We assume that \be\label{e94} n_0^2(x)=1\text{
in }D':=\R^3\setminus D. \ee The function $G(x,y)$ is uniquely
defined (see, e.g., \cite{R509}). The derivations of the results
remain essentially the same because \be\label{e95}
G(x,y)=g_0(x,y)[1+O(|x-y|)],\quad |x-y|\to 0, \ee where
$g_0(x,y)=\frac{1}{4\pi|x-y|}$. Estimates of $G(x,y)$ as $|x-y|\to
0$ and as $|x-y|\to \infty$ are obtained in \cite{R509}. Smallness
of particles in an inhomogeneous medium with refraction coefficient
$n_0^2(x)$ is described by the relation $kn_0a\ll 1$, where
$n_0:=\max_{x\in D}|n_0(x)|$, and $a=\max_{1\leq m\leq M}$diam$D_m$.

\end{document}